\documentclass[a4]{article}

\usepackage{a4wide}
\usepackage{clrscode}
\usepackage{commentstar}
\usepackage{amssymb}
\usepackage{amsmath}
\usepackage[all]{xy}
\usepackage[dvips]{color}
\usepackage{epsfig}
\usepackage{xspace}
\newcommand{\kew}{$k$-edge witness\xspace}
\newcommand{\kvw}{$k$-vertex witness\xspace}

\makeatletter
\DeclareRobustCommand{\qed}{%
  \ifmmode 
  \else \leavevmode\unskip\penalty9999 \hbox{}\nobreak\hfill
  \fi
  \quad\hbox{\qedsymbol}}
\newcommand{\openbox}{\leavevmode
  \hbox to.77778em{%
  \hfil\vrule
  \vbox to.675em{\hrule width.6em\vfil\hrule}%
  \vrule\hfil}}
\newcommand{\qedsymbol}{\openbox}

\newcommand{\proofname}{Proof}
\makeatother

\newtheorem{theorem}{Theorem}

\newtheorem{lemma}{Lemma}

\begin{document}

\title{Worst-case time decremental connectivity and $k$-edge witness}

\author{Andrew Twigg \\ Computing Laboratory \\ Oxford University
  \\ \texttt{andy.twigg@comlab.ox.ac.uk}}

\maketitle

\begin{abstract}
We give a simple algorithm for decremental graph connectivity that handles edge deletions in worst-case time $O(k \log n)$ and connectivity queries in $O(\log k)$, where $k$ is the number of edges deleted so far, and uses worst-case space $O(m^2)$. We use this to give an algorithm for $k$-edge witness (``does the removal of a given set of $k$ edges disconnect two vertices $u,v$?'') with worst-case time $O(k^2 \log n)$ and space $O(k^2 n^2)$. For $k = o(\sqrt{n})$ these improve the worst-case $O(\sqrt{n})$ bound for deletion due to Eppstein et al. We also give a decremental connectivity algorithm using $O(n^2 \log n / \log \log n)$ space, whose time complexity depends on the toughness and independence number of the input graph. Finally, we show how to construct a distributed data structure for \kvw by giving a labeling scheme. This is the first data structure for \kvw that can efficiently distributed without just giving each vertex a copy of the whole structure. Its complexity depends on being able to construct a linear layout with good properties.
\end{abstract}


\section{Introduction}
\label{sec:introduction}

The \kew problem is this: preprocess an undirected graph $G$ so that we can decide if a given set of $k$ edges disconnects two vertices $u,v$, i.e. is the set of edges a witness to the fact that $u,v$ are not $k$-connected? The problem is closely related to decremental graph connectivity, where we support the following operations: $\proc{Delete}(u,v)$, which deletes edge $\{u,v\}$ from the current graph, $\proc{Connected}(u,v)$, which returns \const{true} iff $u,v$ are still connected, and $\proc{Connected}$, which returns \const{true} iff the graph is still connected. The vertex version of the problem is similar except that we instead support $\proc{Delete}(u)$, which deletes a single vertex and all its adjacent edges.

Graph connectivity plays a crucial role in many problems and dynamic connectivity is an intensively studied problem (see \cite{320215} for references). Incremental connectivity (only handling insertions) is equivalent to the union-find problem \cite{545384}. The best bounds for the decremental case are amortized expected -- Henzinger, King and Thorup \cite{320215,280036} used randomization to support updates in $O(\log^2 n)$ expected amortized time, answering queries in $O(\log n / \log \log n)$. In 1998, Holm et al. \cite{502095} obtained a deterministic algorithm with $O(\log^2 n)$ amortized time (this is amortized over adding all the edges to the graph {\em then} deleting them, hence it cannot give good bounds for \kew). The best worst-case bound for decremental connectivity in general graphs is due to Eppstein et al. \cite{EppGalIta-JACM-97} who improved the result of Frederickson \cite{808754} from $O(\sqrt{m})$ to $O(\sqrt{n})$ per update, answering queries in constant time.

In this paper we improve the worst-case bound when the number of deletions is small ($o(\sqrt{n})$), and moreso when the graph has small cutwidth and path covering number, or has a spanning tree with small maximum degree. Our algorithms are simple and reduce the problem of maintaining decremental connectivity to maintaining fully dynamic connectivity (supporting both insertions and deletions) on some auxiliary graph $H$, which usually has size linear in the number of deletions already performed. We can then use known algorithms to maintain connectivity on $H$, and answer queries on $G$ by quickly translating them to queries on $H$. An artefact of our approach is that the time for a deletion depends on the number of vertices or edges already deleted from the original graph, which explains why it only works for small numbers of deletions.

The following is a summary of our results.
\begin{itemize}
\item An algorithm for decremental connectivity using worst-case space $O(m^2)$, handling the $k$th edge deletion in worst-case time $O(k \log n)$ and connectivity queries in $O(k^2)$, where $k$ is the number of edges deleted from $G$ so far. For $k=o(\sqrt{n})$, this improves the deletion time due to Eppstein et al. \cite{EppGalIta-JACM-97}. This gives an an $O(k^2 \log n)$-time algorithm for \kew using space $O(k^2 n^2)$.
\item An alternative algorithm handles deletions in time $O(k^{3/2} + k \log n)$ and connectivity queries in time $O(\log k)$, which improves on Eppstein et al. for $k=o(n^{1/3})$ but has the advantage of small query time.
\end{itemize}
For a graph $G=(V,E)$ let $d_G(v)$ be the degree of vertex $v \in V$, and $\Delta(G) = \max_{v \in V} d_G(v)$. Let $T$ be a spanning tree with minimal $\Delta(T)$ over all spanning trees of $G$.
\begin{itemize}
\item We give a decremental connectivity algorithm using space $O(n^2 \log n / \log \log n)$, handling the $k$th deletion in time $O(\Delta(T)^2 \log n + k \log n)$ and connectivity queries in $O(k^2)$. For Hamiltonian graphs, graphs with bounded independence number, $1/O(1)$-tough graphs and almost all $r$-regular graphs (for fixed $r \geq 3$), this gives $O(k \log n)$ time for deletions.
\end{itemize}
We also construct a {\em distributed} data structure for \kvw using a {\em labeling scheme}: given a graph, assign short labels $L$ to the vertices so that given only $L(u),L(v)$ and $L(S)=L(s_1),\ldots,L(s_k)$, we can efficiently decide if $S$ is a $uv$-cut of size at most $k$ in $G$. We are not aware of any other decremental connectivity algorithm that can be efficiently distributed without giving each vertex a copy of the whole data structure.

In the full version of the paper, we give algorithms for decremental connectivity under vertex deletions with similar times. 

\section{Trees}
As an introduction, we show how to solve \kew and \kvw on $n$-vertex trees in worst-case time $O(k)$. A vertex $w$ is a $uv$-cut in a tree iff the least common ancestor of $u,v$ is an ancestor of $w$ and $w$ is an ancestor of $u$ or $v$. Bender et al. \cite{690192} give a least common ancestor algorithm that answers queries in constant time and uses $O(n)$ space. Hence we can solve \kvw in time $O(k)$ by testing each vertex $s_i$ of the query set $s_1,\ldots,s_k$ and answering `yes' iff any $s_i$ is a $uv$-cut. For \kew, see that an edge $(x,y)$ (where $y$ is the parent of $x$) is a $uv$-cut iff $y$ is a $uv$-cut. To transform this into a decremental connectivity algorithm for trees, store the set of edges deleted so far, and answer each connectivity query in time $O(k)$ by solving \kew.

Interestingly, Bender et al. \cite{690192} use a reduction to the RMQ (range minimum query) problem to solve LCA, which in turn uses the Euler tour data structures. This is quite similar to our connectivity algorithms, which we now describe.

\section{Euler tour}
\label{sec:eulerpaths}

In this section we present an algorithm for decremental graph connectivity based on an Euler tour representation of the graph. An Euler tour of a graph is a tourpath that traverses every edge exactly once. A well-known theorem of Euler states that a graph has an Euler tour iff every vertex has even degree. A simple trick to ensure that $G$ has an Euler tour is to `double up' each edge, so it gets traversed once in each direction.

In an Euler tour $ET(G)$ of $G$, each edge is visited twice (traversed once in each direction) and every degree-$d$ vertex $d$ times. Each time any vertex $u$ is encountered in the tour, we call this an occurence of $u$ and denote the set of occurences of $u$ by $O(u)$, and we shall refer to a particular occurence by its unique position in the tour. If the sequence $ET(G)$ is stored in a balanced binary search tree, then one may insert an interval or splice out an interval (delete an edge of the tour) in time $O(\log n)$, while maintaining the balance of the tree. 

Henzinger and King \cite{320215} use an Euler tour data structure to represent an arbitrary spanning tree of $G$ in their randomized dynamic connectivity algorithm. In our algorithm we construct an Euler tour $ET(G)$ of the whole graph. We maintain an auxiliary graph $H$ where each vertex of $H$ represents a connected interval of $ET(G)$ (i.e. a connected subpath of the Euler tour), and the vertices of $H$ form a disjoint partition of the tour. There is an edge between two vertices of $H$ iff there is some vertex $u$ with an occurence in both intervals. Let $h(u)$ be an arbitrary vertex of $H$ whose interval in $ET(G)$ contains an occurence of $u$. We also maintain a binary search tree on the intervals of $H$ so that in worst-case time $O(\log |H|)$ we can find $h(u)$ for any $u \in G$.

\begin{lemma}
Vertices $u,v \in G$ are connected in $G$ iff $h(u),h(v)$ are connected in $H$.
\end{lemma}
{\bf Proof.}
This follows by the definition of $H$ -- every path from $u$ to $v$ in $G$ corresponds to a set of paths from $h(u)$ to $h(v)$ in $H$, and every path from $u$ to $v$ in $H$ corresponds to a set of paths from $u$ to $v$ in $G$.
\qed

Now we must maintain connectivity on $H$ under vertex insertions, and both edge deletions and insertions. A simple method is to just store its adjacency list representation, then each edge insertion and deletion takes $O(1)$ time, but connectivity queries are answered by using any algorithm for USTCONN on $H$ (e.g. a depth-first search) and take $O(|H|^2)$ time. An alternative is to use the fully-dynamic algorithm of Eppstein which handles edge insertions and deletions in time $O(\sqrt{|H|})$ and answers connectivity queries in time $O(1)$.

\begin{figure}[h]
\begin{center}
\input{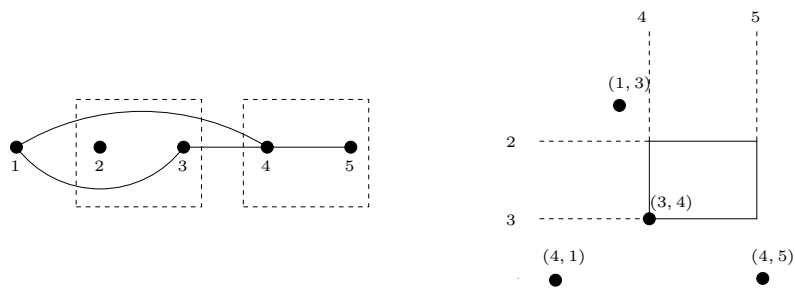}
\end{center}
\label{fig:rangequery}
\caption{The box $(2,4) \times (3,5)$ is empty iff there are no edges between the intervals $[2 \twodots 3]$ and $[4 \twodots 5]$ on the graph}
\end{figure}

The crucial part of our algorithm is the ability to efficiently test for an edge in the graph $H$. We do this by using two-dimensional orthogonal range queries as follows. Given some graph $G$ and a unique identifier $I(u) \in [1 \twodots n]$ for each vertex $u$, an edge $\{u,v\}$ of $G$ is a point $(I(u),I(v))$ on an $n \times n$ grid. Then there is an edge in $G$ with endpoints in the intervals $[a \twodots b],[c \twodots d]$ iff the box $(a,c) \times (b,d)$ is nonempty, as illustrated in Figure \ref{fig:rangequery}.

This applies to our algorithm as follows. For each pair of occurences $u_i,u_j$ of vertex $u$, add a point $(u_i,u_j)$ to the range tree. Now, each vertex of $H$ represents a connected interval $[i \twodots j]$ on the Euler tour. Then two vertices $[a \twodots b],[c \twodots d] \in H$ are adjacent iff there exists some vertex $u$ with occurences in $[a \twodots b]$ and $[c \twodots d]$ iff the box $(a,c) \times (b,d)$ is nonempty. Orthogonal range searching has been extensively studied during the last thirty years with applications to databases and computational geometry. This is the first time we know they have been used for graph connectivity. Chazelle \cite{Chazelle88} has given an algorithm for the static case that stores $r$ points with space $O(r)$ and answers emptiness queries in worst-case time $O(\log r)$ (the algorithm of this section does not remove points).


Now we can describe the algorithm. To delete a edge $\{u,v\}$, the Euler tour is spliced at this edge in time $O(\log n)$, which corresponds to splitting some vertex $h$ of $H$ into two vertices $h_1,h_2$. The edges of $h_1$ union the edges of $h_2$ are equal to the edges of $h$, so we need only test each of the neighbours of $h$ with $h_1$ and $h_2$ to see if there should be an edge. By definition of $H$, there is an edge between two vertices of $H$ iff their corresponding intervals on the Euler tour contain an occurence of the same vertex of $G$. This test is done in worst-case time $O(\log n)$ for each edge by asking the range tree if some box is empty. A connectivity query for $u,v$ is handled by finding $h(u),h(v)$ and then calling $\proc{Connected}_H(h(u),h(v))$. To handle the query $\proc{Connected}$, we simply call $\proc{Connected}_H$ since $G$ is connected iff $H$ is connected.

\begin{codebox}
\Procname{$\proc{Initialise}(G)$}
\li Double up each edge of $G$
\li $ET(G) \gets $ an Euler tour of $G$
\li \For each vertex $u$
\li   \Do $O(u) \gets$ the set of occurences of $u$ in $ET(G)$
\li       \For each pair $u_i,u_j \in O(u)$
\li         \Do $\proc{Range-Tree-Insert}(u_i,u_j)$
            \End
      \End
\li Let $H$ be a graph with a single vertex $[1 \twodots 2m]$.
\end{codebox}

\begin{codebox}
\Procname{$\proc{Delete}(u,v)$}
\li \For each occurence $i$ of edge $\{u,v\}$ in $ET(G)$
\li \Do $\proc{Euler-Tour-Delete}(i)$ \RComment $O(\log n)$
\li     $[a \twodots b] \gets h(i)$ \RComment $O(\log |H|)$
\li     $\proc{Insert-Edge}_H([a \twodots i])$ \RComment $O(1)$
\li     $\proc{Insert-Edge}_H([i+1 \twodots b])$
\li     \For each neighbour $[c \twodots d]$ of $[a \twodots b]$ in $H$
\li         \Do \If $\proc{Box-Not-Empty}((a,c),(i,d))$ \RComment $O(\log n)$
\li                \Then $\proc{Insert-Edge}_H([a \twodots i],[c \twodots d])$ \RComment $O(1)$
                \End
\li             \If $\proc{Box-Not-Empty}((i+1,c),(b,d))$
\li                \Then $\proc{Insert-Edge}_H([i+1 \twodots b], [c \twodots d])$
                \End
             \End
\li    $\proc{Delete-Vertex}_H([a \twodots i])$ \RComment $O(\Delta(H))$
    \End
\end{codebox}

\begin{codebox}
\Procname{$\proc{Connected}(u,v)$}
\zi \Comment Returns true iff $u,v$ are connected in $G$
\li $[a \twodots b] \gets h(u)$ \RComment $O(\log |H|)$
\li $[c \twodots d] \gets h(v)$
\li \Return $\proc{Connected}_H([a \twodots b], [c \twodots d])$ \RComment $O(|H|^2)$
\end{codebox}

\begin{codebox}
\Procname{$\proc{Connected}$}
\zi \Comment Returns true iff $G$ is connected
\li \Return $\proc{Connected}_H$ \RComment $O(|H|^2)$
\end{codebox}

\begin{codebox}
\Procname{$k$-\proc{Edge-Witness}$(u,v,\{x_1,y_1\},\ldots,\{x_k,y_k\})$}
\li $G' \gets$ a sparse $k$-connectivity certificate of $G$
\li $\proc{Initialise}(G')$
\li Let $H$ contain a single vertex $[1 \twodots 2m]$
\li \For each $i \in \{1,\ldots,k\}$
\li     \Do $\proc{Delete}(\{x_i,y_i\})$
        \End
\li \Return $\proc{Connected}(u,v)$
\end{codebox}

\subsection{Complexity}

After $k$ edges have been deleted from $G$, $H$ has at most $k+1$ vertices. Each edge occurs twice in the euler tour, so the loop at line 1 of $\proc{Delete}$ only incurs a constant factor. It takes time $O(\log n)$ to splice out an interval of the euler tour, and each range tree query takes time $O(\log n)$. If we use the adjacency list to represent $H$, the times are as shown in the code. The loop at line 6 is repeated $O(\Delta(H))$ times. Hence procedure $k$-\proc{Edge-Witness} takes worst-case time $O(k^2 \log n)$. Additionally, in time $O(k^2)$ we can construct the reachability matrix for all vertices of $H$ to subsequently test if the same set of edges is a witness for the non-$k$-connectedness of any pair of vertices in $G$, in time $O(\log k)$.

Alternatively, using the fully-dynamic connectivity algorithm of Eppstein et al. makes lines 4,5,8,9 take time $O(\sqrt{|H|})$ and line 11 takes time $O(\Delta(H) \sqrt{|H|})$, so $\proc{Delete}$ takes worst-case time $O(k^{3/2} + k \log n)$, but $\proc{Connected}(u,v)$ and $\proc{Connected}$ take $O(\log k)$ time.

The space requirement is dominated by the cost of storing the vertex occurences in the range tree, using $O(\sum_{v \in G} d_G(v)^2)=O(m^2)$ bits. If we are only interested in cuts of size at most $k$ (as for \kew), we can use less space. A {\em sparse $k$-connectivity certificate} for a graph $G$ is a subgraph $G'$ containing at most $kn$ edges, such that any cut of value at most $k$ in $G$ has the same value in the certificate. Nagamochi and Ibaraki \cite{NagamochiI92} give a linear-time algorithm for computing sparse $k$-connectivity certificates.

The modified algorithm is the same except that we replace $G$ with its sparse $k$-connectivity certificate in $\proc{Initialise}$. The correctness follows from the properties of the connectivity certificate. The space complexity is reduced to that needed to store the vertex occurences in the range tree for the Euler tour of the certificate, i.e. $O(\sum_{v \in V} d_{G'}(v)^2) = O(k^2 n^2)$ bits in the worst case that some vertex has almost all the edges. If the certificate has degree at most $k$ then the space requirement is reduced to $O(k^2 n)$ bits. The query time is unchanged, as it only depended on the set of edges being deleted.

\section{An algorithm using less space}
\label{sec:treelayout}

In this section we give an decremental algorithm that uses space $O(n^2 \log n / \log \log n)$ and handles the $k$th deletion in time $O(\Delta(T)^2 \log n + k \log n)$. It answers connectivity queries in time $O(k^2)$, and therefore gives an $O(k \Delta(T)^2 \log n + k^2 \log n)$-time algorithm for \kew, which is $O(k^2 \log n)$ for Hamiltonian graphs, graphs of bounded independence number, $1/O(1)$-tough graphs and almost all $r$-regular graphs (for fixed $r \geq 3$). As in the previous section, we can also use a different algorithm to maintain connectivity on the auxiliary graph and this gives $O(\Delta(T)^2 \log n + k^{3/2} \log n)$ time for deletions but with $O(\log k)$ query time.

The algorithm is most similar to that of Henzinger and King \cite{320215} with the main exception that instead of {\em maintaining} a spanning forest of $G$, we do not bother to patch up tree edges when they are removed. Instead, we keep track of the fragmented parts of the forest as edges are removed (using a dynamic connectivity algorithm) and use this information to efficiently answer the connectivity queries. Consider a graph $G=(V,E)$ and a spanning tree $T=(V,F)$. We construct the Euler tour $ET(T)$ on $T$ with its edges doubled up by calling the following procedure with the root vertex. The sequence produced can be stored in a binary search tree so that tree edge deletions (interval splicing) take time $O(\log n)$.
\begin{codebox}
\Procname{$\proc{ET}(v)$}
\li visit $v$
\li \For each child $u$ of $v$
\li     \Do $\proc{ET}(u)$
\li         visit $u$
        \End
\end{codebox}
We maintain a graph $H$ whose vertices are intervals on the Euler tour $ET(T)$ of the spanning tree $T$, and build the range tree with the points representing pairs of vertex occurences in $ET(T)$. The number of points is at most $\sum_{v \in T} d_T(v)^2 = O(n \min(n, \Delta(T)^2))$ since $T$ is a tree and so $\sum_{v \in T} d_T(v) = n-1$. However, we must also handle the nontree edges of $G$. We do this by adding to the range tree, for each {\em nontree} edge $\{u,v\} \in G$, a point $(u_i,v_j)$ for each pair of occurences $u_i \in O(u), v_j \in O(v)$. Since each vertex appears in the Euler tour at most $\Delta(T)$ times, this is $$\sum_{\{u,v\} \in E \setminus F} d_T(u) d_T(v) \leq \sum_{ \{u,v\} \in E } d_T(u) d_T(v) = \left( \sum_u d_T(u) \right)^2 = \left( 2n - 2 \right)^2$$ for any spanning tree $T=(V,F)$ of $G=(V,E)$. This dominates the space used to store the vertex occurences and so we use at most $O(n^2)$ points in total.

Assume that we represent the auxiliary graph $H$ with its adjacency list (so that edge operations are $O(1)$). Deleting a tree edge is handled as before: we delete the two appearences of the edge from the Euler tour of $T$, which splits some vertex of $H$ into two vertices, and then test the new edges of the new vertices of $H$ using emptiness queries. This takes total time $O(k \log n)$, after $k$ edges have been deleted. Deleting a nontree edge $\{u,v\}$ has no effect on the Euler tour but now we must delete all the points $(u_i,v_j) \in O(u) \times O(v)$ from the range tree, corresponding to the edge $\{u,v\}$. Mortensen \cite{644210} has given a dynamic range query data structure which handles emptiness queries and deletions in worst-case time $O(\log r)$ and uses space $O(r \log r / \log \log r)$ for $r$ points. Therefore our algorithm handles the $k$th deletion in worst-case time $O(\Delta(T)^2 \log n + k \log n)$ and uses space $O(n^2 \log n / \log \log n)$.


\subsection{Complexity}

The algorithm relies on being able to construct a spanning tree of $G$ with small maximum degree, or more accurately, if $\{u,v\}$ is a {\em nontree} edge of $G$ then the product $d_T(u) d_T(v)$ should be small, i.e. we want all the high degree vertices of $T$ to be neighbours in $T$ also.

Let $\Delta^*(G,T)$ be the smallest integer $d$ such that $G$ has a spanning tree of maximum degree $d$. It is NP-hard to determine exactly $\Delta^*(G,T)$ since $\Delta^*(G,T)=2$ iff $G$ has a Hamiltonian path, which is NP-complete \cite{574848}. Furer and Raghavachari \cite{furer94} give a polynomial-time algorithm that outputs a spanning tree $T$ with degree at most $\Delta^*(G,T) + 1$. A theorem of Dirac \cite{dirac52} says that if each vertex of $G$ has degree at least $n/2$ then $G$ contains a Hamiltonian cycle. A graph $G=(V,E)$ is $t$-tough if the number of connected components of $G \setminus S$ is at most $|S| / t$ for every cutset $S \subseteq V$. In 1989, Win proved the following theorem.
\begin{theorem}[\cite{win89}]
Let $t$ be a positive integer. Every $1/t$-tough graph $G$ has a spanning tree of degree $t+2$, i.e. $\Delta^*(G,T) = O(t)$.
\end{theorem}
Combining the above theorem with the algorithm we get the following result.
\begin{theorem}
Let $G$ be $1/t$-tough. There is a decremental connectivity algorithm with polynomial preprocessing time for $G$ where the $k$th deletion takes time $O(t^2 \log n + k \log n)$ and connectivity queries $O(k^2)$, using space $O(n^2 \log n / \log \log n)$. Also, \kew can be solved in time $O(k t^2 \log n + k^2 \log n)$.
\end{theorem}

\section{A distributed data structure for \kvw}
\label{sec:labelingscheme}

In this section we show how to construct a distributed data structure for \kvw by giving a labeling scheme for the problem. This consists of a marker algorithm that takes the graph and outputs a binary string called a {\em label} $L(u)$ for each vertex $u$. There is also a decoder algorithm that, given only $L(u),L(v)$ and $L(S)=L(s_1), \ldots, L(s_k)$, decides if the set $S$ of vertices is a $uv$-cut in $G$. We give only a brief outline here, and defer a more detailed exposition to the full version of the paper.

We consider two variants of the problem: in the first, each vertex $u$ has a set $S(u)$, which is given to the marker algorithm, and the decoder takes $L(u),L(v)$ and decides if $S(u)$ is a $uv$-cut in $G$. In the second variant, the marker algorithm only has knowledge of the graph $G$, and the decoder takes $L(u),L(v)$ and $\{L(s_i)\}$ for each $s_i \in S$, and decides if $S$ is a $uv$-cut in $G$.

\subsection{The sets are given to the marker algorithm}

When the sets are given to the marker algorithm, the problem becomes how to efficiently represent the sets, and the partitions they induce in $G$ (if any). For the case where $S$ is a set of edges, we can use the tree layout algorithm of Section \ref{sec:treelayout} to compute an auxiliary graph $H_u$ corresponding to each set $S(u)$. The label $L(u)$ then stores the following: $u$'s its identifier in the tree (from the postorder traversal), a mapping $g_u: \{1,\ldots,n\} \to \{0,1\}$ from vertex ids in the tree to binary values, where $g_u(v)=0$ iff the vertex of $H_u$ containing $v$ can be reached from the vertex of $H_u$ containing $u$, i.e. iff the set $S(u)$ is a $uv$-cut in $G$. This gives a labeling scheme where the label $L(u)$ is of size $O(|H_u|)=O(|S(u)|)$ bits, since the mapping $g_u$ can be stored using a binary search tree on the intervals of $H_u$ (where the key is the smallest identifier contained in the interval). Given two labels $L(u),L(v)$, the decoder algorithm looks up the identifier of $v$ in $g_u$ and returns the binary value stored there, in time $O(\log |S(u)|)=O(\log n)$. If $S$ is a set of vertices, we can use the same method but with $H(u)$ having $O({\mathrm degree}(G) |S|)$ vertices.

\subsection{The sets are not given to the marker algorithm}

The algorithm in Section \ref{sec:eulerpaths} used an Euler path to efficiently represent the edge cuts of a graph. Here we use a path cover of $G$ to efficiently represent the vertex cuts of a graph, and because we shall be able to efficiently distribute it. A family of paths $P_1,\ldots,P_k$ in a graph $G$ is a {\em path cover} of $G$ if every vertex of $G$ is contained in exactly one of these paths. The {\em path covering number} $\vartheta(G)$ of $G$ is the minimum number of paths in a path cover of $G$. Computing $\vartheta(G)$ exactly is NP-hard since $\vartheta(G)=1$ iff $G$ contains an Hamiltonian path.

A linear layout of a graph with $n$ vertices is a bijection $\mathcal{L} : V \to \{1,\ldots,n\}$. For a vertex $v$, let $\mathrm{cutwidth}(\mathcal{L},u)$ be the number of edges crossing the point $\mathcal{L}(u)$ in the layout. The cutwidth of a layout $\mathrm{cutwidth}(\mathcal{L})$ is $\max_u \mathrm{cutwidth}(\mathcal{L},u)$, i.e. the maximum number of edges crossing any point in the layout. The cutwidth of a graph $G$ is the minimum cutwidth of a layout of $G$. Deciding whether $\mathrm{cutwidth}(G) \leq k$ (for an integer $k$) is NP-complete \cite{574848}, but is approximable to within a factor of $O(\log n \log \log n)$ in polynomial time \cite{347478}. For a linear layout $\mathcal{L}$ of $G$, we say that there is a {\em hole} at position $i$ in the layout if there is no edge $\{ \mathcal{L}^{-1}(i),\mathcal{L}^{-1}(i+1) \}$ in $G$, and let $\mathrm{holes}(\mathcal{L})$ be the set of holes in the layout. We have the following observation.

\begin{lemma}
\label{lem:linearlayout1}
Any graph $G$ has a linear layout with $\vartheta(G)-1$ holes.
\end{lemma}
{\bf Proof.}
Given a graph $G$, we can construct a linear layout with $\vartheta(G)-1$ holes: for each path $P=v_1,\ldots,v_k$ in the path cover, add the vertices of the path to the layout in the order they appear on the path. The layout clearly has the claimed number of holes.

To see that every linear layout with $h$ holes gives a path cover of size $h+1$, simply remove all the crossing edges (those not on the line). The spine of the layout then consists of $h+1$ vertex-disjoint paths where each vertex of $G$ appears in exactly one path, and the paths cover $G$. \qed

The algorithm works as follows. Construct a linear layout having both small cutwidth {\em and} a small number of holes (we note that we do not currently know of any linear layout algorithm that simultaneously minimizes both these quantities, and moreso, they appear to be opposing so the best we might hope for is some sort of tradeoff -- we discuss this in the full paper). We maintain incremental connectivity (edge insertions only) on an auxiliary graph $H$ where each vertex of $H$ represents an interval with no holes on the layout, with an edge between two intervals if there is an edge in $G$ with endpoints in both intervals. However, we shall not use the range tree.

We must also deal with the fact that intervals of the layout do not necessarily represent connected subgraphs of $G$, due to the holes. We will handle the holes in the following simple way: for a hole at position $i$, we pretend that there is a dummy vertex $v$ connected to both $i$ and $i+1$, then proceed as before except that the query set is now $S \cup \{v\}$. So we add the dummy vertex $v$ and then use our decremental connectivity algorithm to delete it at query time. The correctness of the algorithm is a consequence of the following simple lemma.

\begin{lemma}
\label{lem:linearlayout2}
Vertices $u,v$ are connected in $G$ iff $h(u),h(v)$ are connected in $H$.
\end{lemma}
{\bf Proof.}
This follows by the definition of $H$: it is a disjoint partition of the vertices of $G$ into connected subgraphs with an edge between two subgraphs in $H$ iff there is an edge between two vertices in the subgraphs in $G$. Therefore, every path from $u$ to $v$ in $G$ corresponds to exactly one path from $h(u)$ to $h(v)$ in $H$, and every path from $u$ to $v$ in $H$ corresponds to a set of paths from $u$ to $v$ in $G$.
\qed

Instead of storing the set of crossing edges centrally, the label for vertex $u$ stores a list of its crossing edges in the layout. To handle the holes, each vertex can simply store the {\em labels} for all the holes. However, the following lemma shows that we need only consider a subset of these holes. For a graph $G$, a set $S$ and two vertices $u,v$ with $S$ being a $uv$-cut in $G$, a hole is {\em important} if replacing it with an edge would result in $u,v$ being connected in $G \setminus S$. We can prove the following lemma.

\begin{lemma}
\label{lem:importantholes}
For a linear layout $\mathcal{L}$, let $\mathrm{holes}(\mathcal{L},u)$ be the set of holes spanned by the crossing edges of vertex $u$ in the layout. Then all the important holes for $S,u,v$ are covered by $\mathrm{holes}(\mathcal{L},u) \cup \mathrm{holes}(\mathcal{L},v) \cup \bigcup_{i=1}^k \mathrm{holes}(\mathcal{L},s_i)$.
\end{lemma}

The decoder algorithm is given below. We create all the vertices of $H$ and store their intervals in a binary tree (so that we can find the interval $h(u)$ containing vertex $u$ in time $O(\log |H|)$), then use the \proc{Union-Find} data structure \cite{545384} to maintain incremental connectivity as we add the edges. This allows us to implement \proc{Find} in worst-case time $O(\log |H|)$ and \proc{Union} in worst-case time $O(1)$ (these worst-case bounds are known to be optimal). The inner loop at line 7 is executed at most $O(|H| \mathrm{cutwidth}(\mathcal{L}))=O((k+|\mathrm{holes}(\mathcal{L})|) \mathrm{cutwidth}(\mathcal{L}))$ times, so the decoder takes worst-case time $O((k+|\mathrm{holes}(\mathcal{L})|) \mathrm{cutwidth}(\mathcal{L}) \log n)$.

\begin{codebox}
\Procname{$k$-\proc{Vertex-Witness-Decoder}$(L(u),L(v),L(s_1),...,L(s_k))$}
\li $T \gets \mathrm{holes}(\mathcal{L})$
\li sort the elements of $S \cup T$ by their position in the layout 
\li $H \gets$ set of intervals of $G \setminus (S \cup T)$
\li Construct the binary search tree on intervals of $H$
\li \For each vertex $k \in H$
\li     \Do Insert $k$ into the $\proc{Union-Find}$ structure \End \RComment $O(1)$
\li \For each vertex $k \in (S \cup T)$
\li     \Do \For each crossing edge $(i,j)$ of $k$
\li             \Do $x \gets \proc{Find}(h(i))$ \RComment $O(\log |S \cup T|)$
\li                 $y \gets \proc{Find}(h(j))$
\li                 $\proc{Union}(x, y)$ \RComment $O(1)$
                \End
        \End
\li \Return $\proc{Find}(h(u)) \neq \proc{Find}(h(v))$ \RComment $O(\log |S \cup T|)$
\end{codebox}


The label assigned to a vertex $u$ is has worst-case size $O(|\mathrm{holes}(\mathcal{L},u)| \mathrm{cutwidth}(\mathcal{L}) \log n)$. We note that a consequence of using the \proc{Union-Find} data structure is that after the decoder algorithm has finished, we can repeatedly answer connectivity queries in worst-case time $O(\log n)$ on $G \setminus S$.

It is worth noting that any labeling scheme with labels of size $O(s)$ bits gives a nondistributed data structure of size $O(n s)$ bits with the same time complexities. Using the above labeling scheme, and storing the crossing edges of each hole exactly once gives a nondistributed data structure using $O(\sum_{u \in G} \mathrm{cutwidth}(\mathcal{L},u) \log n) = O(\mathrm{wirelength}(\mathcal{L}) \log n)$ bits total space, handling queries in the same time as above.


We can prove the following simple lower bound on the label size.
\begin{theorem}
\label{thm:lowerbound}
Any labeling scheme for the \kvw problem on $n$-vertex graphs must assign to some vertex a label of $\Omega(n)$ bits in the worst-case.
\end{theorem}
{\bf Proof.}
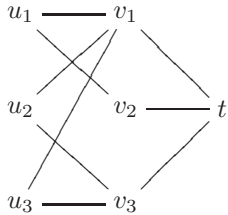
\begin{figure}
  \caption{Illustrating Theorem \ref{thm:lowerbound}}
\[
\vcenter{
\xymatrix{
u_1 \ar@{-}[r] \ar@{-}[dr] & v_1 \ar@{-}[dr] & \\
u_2 \ar@{-}[ur] \ar@{-}[dr] & v_2 \ar@{-}[r] & t \\
u_3 \ar@{-}[r] \ar@{-}[uur] \ar@{-}[r] & v_3 \ar@{-}[ur] &  \\
}}
\]
\label{fig:lowerbound}
\end{figure}
By reduction from adjacency on bipartite graphs. Given a bipartite graph $G=((U,V),E)$, construct $G'$ by adding a vertex $t$, connected to each vertex of $V$, as in Figure \ref{fig:lowerbound}. Now consider two vertices $u \in U$ and $v \in V$. It is clear that $u$ is not adjacent to $v$ in $G$ iff $V \setminus \{v\}$ is a cut between $u$ and $t$ in $G'$. Since any labeling scheme for deciding adjacency on bipartite graphs must assign some vertex a label of size $\Omega(n)$ bits, the theorem follows. \qed



\section{Decremental connectivity on small cutwidth graphs}

Assume that we can construct a linear layout of $G$ with bounded cutwidth and a bounded number of holes. In this section we show that decremental connectivity can be done in time $O(\sqrt{k})$ for the $k$th deletion, with connectivity queries answered in time $O(\log k)$. This lets us answer \kvw and \kew in worst-case time $O(k)$. The algorithm is essentially the same as in the previous section, but we make use of the following observation.
\begin{lemma}
The degree of any vertex of $H$ is at most twice the cutwidth of the layout.
\end{lemma}
{\bf Proof.}
Consider some vertex $h \in H$ representing an interval with vertices $u,v$ being the endpoints of this interval. Every edge of $h$ corresponds to a crossing edge of either $u$ or $v$ in the linear layout of $G$, i.e. leaving the interval to the left or to the right.
\qed

The algorithm is given below. On the linear layout, we call an interval $[i \twodots j]$ a {\em maximal connected interval} if there are no holes in $[i \twodots j]$ and every interval containing $[i \twodots j]$ also contains a hole. It is easy to see that the set of maximal connected intervals gives a unique partition of the layout into disjoint intervals.

Initially, we add all the maximal intervals to the auxiliary graph $H$ and the crossing edges of the layout to the static range tree (for example, using the algorithm of Chazelle \cite{Chazelle88}). We then use the fully dynamic algorithm of Eppstein et al. \cite{EppGalIta-JACM-97} to maintain connectivity on $H$, at a cost of $O(\sqrt{|H|})$ per edge insertion or deletion.

\begin{codebox}
\Procname{$\proc{Initialise}$}
\li $\mathcal{L} \gets$ a linear layout of $G$
\li \For each edge $\{u,v\} \in G$
\li     \Do $\proc{Range-Tree-Insert}(L(u),L(v))$
        \End
\li \For each maximal connected interval $[i \twodots j]$ on $\mathcal{L}$
\li     \Do Add the vertex $[i \twodots j]$ to $H$
        \End
\li \For each pair of vertices $[a \twodots b], [c \twodots d]$ in $H$
\li     \Do \If $\proc{Box-Not-Empty}((a,c),(b,d))$
\li             \Then Add edge $([a \twodots b], [c \twodots d])$ to $H$
            \End
    \End
\end{codebox}

\begin{codebox}
\Procname{$\proc{Delete}(u)$}
\li $[a \twodots b] \gets h(u)$ \RComment $O(\log |H|)$
\li \If $(a = b)$
\li     \Then \Return
    \End
\li \Comment Otherwise, split $[a \twodots b]$ at $\mathcal{L}(u)$
\li $\proc{Insert-Vertex}_H([a \twodots \mathcal{L}(u)-1])$
\li $\proc{Insert-Vertex}_H([\mathcal{L}(u)+1 \twodots b])$ \RComment $O(\sqrt{|H|})$

\li \For each neighbour $[c \twodots d]$ of $h(u) \in H$
\li     \Do \If $\proc{Box-Not-Empty}((a,c),(\mathcal{L}(u)-1,d))$ \RComment $O(\log n)$
\li             \Then $\proc{Insert-Edge}_H(([a \twodots I(u)-1], [c \twodots d])$ \RComment $O(\sqrt{|H|})$
            \End
\li         \If $\proc{Box-Not-Empty}((\mathcal{L}+1,a),(b,d))$
\li             \Then $\proc{Insert-Edge}_H([\mathcal{L}+1 \twodots b],[c \twodots d])$
            \End
        \End
\li $\proc{Delete-Vertex}_H([a \twodots b])$ \RComment $O(\sqrt{|H|} \Delta(H))$
\end{codebox}


\begin{codebox}
\Procname{$\proc{Connected}(u,v)$}
\li \Return $\proc{Connected-in-H}(h(u),h(v))$ \RComment $O(\log |H|) + O(1)$
\end{codebox}

\subsection{Analysis}

Let the layout have $j$ holes. Since $|H| \leq k + j$ and $\Delta(H) \leq 2 \mathrm{cutwidth}(\mathcal{L})$, the $k$th deletion takes worst-case time $O(\sqrt{k+j} \min(k+j,\mathrm{cutwidth}(\mathcal{L})))$ and connectivity queries are answered in $O(\log k)$. Alternatively, we could just store $H$ using adjacency lists, then the $k$th update takes worst-case time $O(\min(k+j,\mathrm{cutwidth}(\mathcal{L})))$ but queries take time $O((k+j)^2)=O(k^2 + j^2)$ by performing a depth-first search of the auxiliary graph $H$.

A modification of the above algorithm is to only add the holes as they are required, rather than adding them all in the initialisation step. If the queries do not involve vertices whose crossing edges cover many holes, then this may be more efficient than the previous algorithm.

{\bf Remark.}
We could use the linear layout to handle edge deletions by deleting the edges from the range tree, but we would still incur the same worst-case cost as a single vertex deletion.

\section{Further work}

There are many interesting open problems in dynamic connectivity. Probably the most important is still to break the $O(\sqrt{n})$ worst-case bound for a single deletion, over an {\em arbitrarily long} sequence of deletions. Very recently, Thorup \cite{1060607} gave a fully-dynamic all-pairs shortest paths algorithm with good worst-case time by constructing a data structure that supports any sequence (called an {\em epoch}) of $\Delta$ deletions in $\tilde O(n^{2.5} \sqrt{\Delta})$ total time, then partially rebuilding the structure in the background to avoid incurring a single expensive update. We have tried to apply a similar technique to our algorithms but with no luck -- although we can support an epoch of $O(\sqrt{n})$ deletions in total time $\tilde O(n)$ with worst-case $\tilde O(\sqrt{n})$ per deletion, which allows us to spend at most an extra $\tilde O(\sqrt{n})$ time per deletion on rebuilding the data structure, this only amounts to $\tilde O(n)$ time over the epoch. To break the $o(\sqrt{n})$ bound for deletions, we would have to have $o(\sqrt{n})$-length epochs, but then we would only be able to spend $o(n)$ rebuilding time during each epoch.

It would also be interesting to investigate other ways of constructing distributed data structures for important graph connectivity problems such as \kvw, as these would appear to be very useful in large distributed networks such as the Internet.



{\bf Acknowledgments}
We wish to thank Andrew Thomason for a very helpful observation.

\small
\bibliographystyle{acm}





\newpage
\appendix
\section{Proof of Lemma \ref{lem:linearlayout1}}

{\bf Proof.}
Given a graph $G$, we can construct a linear layout with $\vartheta(G)-1$ holes: for each path $P=v_1,\ldots,v_k$ in the path cover, add the vertices of the path to the layout in the order they appear on the path. The layout clearly has the claimed number of holes.

To see that every linear layout with $h$ holes gives a path cover of size $h+1$, simply remove all the crossing edges (those not on the line). The spine of the layout then consists of $h+1$ vertex-disjoint paths where each vertex of $G$ appears in exactly one path, and the paths cover $G$. \qed

\section{Proof of Lemma \ref{lem:linearlayout2}}

{\bf Proof.}
This follows by the definition of $H$: it is a disjoint partition of the vertices of $G$ into connected subgraphs with an edge between two subgraphs in $H$ iff there is an edge between two vertices in the subgraphs in $G$. Therefore, every path from $u$ to $v$ in $G$ corresponds to exactly one path from $h(u)$ to $h(v)$ in $H$, and every path from $u$ to $v$ in $H$ corresponds to a set of paths from $u$ to $v$ in $G$.
\qed




\section{Proof of Theorem \ref{thm:lowerbound}}

{\bf Proof.}
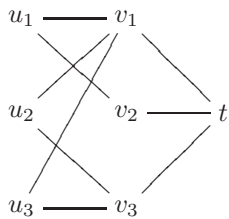
\begin{figure}
  \caption{Illustrating Theorem \ref{thm:lowerbound}}
\[
\vcenter{
\xymatrix{
u_1 \ar@{-}[r] \ar@{-}[dr] & v_1 \ar@{-}[dr] & \\
u_2 \ar@{-}[ur] \ar@{-}[dr] & v_2 \ar@{-}[r] & t \\
u_3 \ar@{-}[r] \ar@{-}[uur] \ar@{-}[r] & v_3 \ar@{-}[ur] &  \\
}}
\]
\label{fig:lowerbound}
\end{figure}
By reduction from adjacency on bipartite graphs. Given a bipartite graph $G=((U,V),E)$, construct $G'$ by adding a vertex $t$, connected to each vertex of $V$, as in Figure \ref{fig:lowerbound}. Now consider two vertices $u \in U$ and $v \in V$. It is clear that $u$ is not adjacent to $v$ in $G$ iff $V \setminus \{v\}$ is a cut between $u$ and $t$ in $G'$. Since any labeling scheme for deciding adjacency on $G$ must assign some vertex a label of size $\Omega(n)$ bits, the theorem follows. \qed

\end{document}